\documentclass[a4paper,latin9]{aa}
\usepackage[latin9]{inputenc}
\usepackage{amsmath}
\usepackage{amssymb}
\usepackage{graphicx}

\makeatletter

\providecommand{\tabularnewline}{\\}

\usepackage{times}\usepackage{multirow}\usepackage{rotating}\usepackage{longtable}\usepackage{lscape}
\usepackage{lipsum}
\usepackage{times}
\usepackage{multirow}
\usepackage{rotating}
\usepackage{longtable}
\usepackage{lscape}\usepackage{epstopdf}
\defcitealias{pierre15}{Paper I}
\defcitealias{pacaud15}{Paper II}
\defcitealias{giles15}{Paper III}
\defcitealias{lieu15}{Paper IV}
\defcitealias{fotopoulou15}{Paper VI}
\defcitealias{pompei15}{Paper VII}
\defcitealias{lidman15}{Paper XIV}
\defcitealias{koulouridis15}{Paper XII}
\defcitealias{civano_inprep}{submitted}

\def\f#1   {Fig.~\ref{#1}}
\def\s#1   {Sec.~\ref{#1}}
\def\tab#1   {Tab.~\ref{#1}}
\def\eq#1   {Eq.~\ref{#1}}
\def\t#1   {Tab.~\ref{#1}}

\def\comm#1   {{\tt (COMMENT: #1) }}

\titlerunning{ATCA XXL-S Pilot Project}
\authorrunning{Smolcic et al.}

\makeatother

\begin{document}

\title{The XXL Survey: XI. ATCA 2.1 GHz continuum observations}

\author{Vernesa Smol{\v{c}}i{\'{c}}\inst{1}, Jacinta Delhaize\inst{1},
Minh Huynh\inst{2}, Marco Bondi\inst{3}, Paolo Ciliegi\inst{4},
Mladen Novak\inst{1}, Nikola~Baran\inst{1}, Mark~Birkinshaw\inst{5},
Malcolm N. Bremer$^{5}$, Lucio Chiappetti\inst{6}, Chiara Ferrari$^{7}$,
Sotiria Fotopoulou$^{8}$, Cathy Horellou$^{9}$, Sean L. McGee$^{10}$,
Florian~Pacaud$^{11}$, Marguerite Pierre$^{12}$, Somak Raychaudhury$^{10}$,
Huub Röttgering$^{13}$, Cristian Vignali$^{14,4}$ }

\institute{University of Zagreb, Physics Department, Bijeni\v{c}ka cesta 32,
10002 Zagreb, Croatia \\ e-mail: \url{vs@phy.hr}
 \and International Centre for Radio Astronomy
Research (ICRAR), M468, University of Western Australia, 35 Stirling
Hwy, WA 6009, Australia \and Istituto di Radioastronomia di Bologna
- INAF, via P. Gobetti, 101, 40129, Bologna, Italy \and INAF-Osservatorio
Astronomico di Bologna, Via Ranzani 1, I - 40127 Bologna, Italy \and
H.H. Wills Physics Laboratory, University of Bristol, Tyndall Avenue, Bristol BS8 1TL, U.K.
\and INAF, IASF Milano, via Bassini 15, I-20133 Milano, Italy \and Laboratoire Lagrange, Université Côte d'Azur, Observatoire de la Côte d'Azur, CNRS, 
Blvd de l'Observatoire, CS 34229, 06304 Nice cedex 4, France  \and Department
of Astronomy, University of Geneva, ch. d'Ecogia 16, CH-1290 Versoix,
Switzerland \and Dept. of Earth \& Space Sciences, Chalmers University
of Technology, Onsala Space Observatory, SE-439 92 Onsala, Sweden
\and School of Physics and Astronomy, University of Birmingham, Edgbaston,
Birmingham B15 2TT, UK \and Argelander Institut für Astronomie, Universität
Bonn, D-53121, Bonn \and Service d'Astrophysique AIM, CEA/DSM/IRFU/SAp, CEA Saclay,
F-91191 Gif sur Yvette \and Leiden Observatory, Leiden University,
PO Box 9513, NL-2300 RA Leiden, the Netherlands \and Dipartimento
di Fisica e Astronomia Universita degli Studi di Bologna, Viale Berti
Pichat 6/2, 40127, Bologna, Italy }

\abstract{We present 2.1\,GHz imaging with the Australia Telescope Compact
Array (ATCA) of a 6.5 deg$^{2}$ region within the XXM-Newton XXL
South field using a band of $1.1-3.1$\,GHz. We achieve an angular
resolution of $4.7''\times4.2''$ in the final radio continuum map
with a median rms noise level of 50\,$\mu$Jy/beam. We identify 1389
radio sources in the field with peak S/N$\geq5$ and present the catalogue
of observed parameters. We find that 305 sources are resolved, of
which 77 consist of multiple radio components. These number counts
are in agreement with those found for the COSMOS-VLA 1.4\,GHz survey.
We derive spectral indices by a comparison with the Sydney University Molongolo Sky Survey (SUMSS) 843\,MHz data.
We find an average spectral index of -0.78 and a scatter of 0.28,
in line with expectations. This pilot survey was conducted in preparation
for a larger ATCA program to observe the full 25\,deg$^{2}$ southern
XXL field. When complete, the survey will provide a unique resource of sensitive,
wide-field radio continuum imaging with complementary X-ray data in
the field. This will facilitate studies of the physical mechanisms
of radio-loud and radio-quiet AGNs and galaxy clusters, and the role
they play in galaxy evolution. The source catalogue is publicly
available online via the XXL Master Catalogue browser and the Centre de Donn\'ees astronomiques de Strasbourg (CDS).}


\maketitle
\makeatother

\section{Introduction\label{sec:intro}}

\noindent Low-luminosity radio-loud AGN ($L_{1.4\,{\rm GHz}}\lesssim10^{25}$\,W/Hz)
are somewhat puzzling systems which do not fit into the Unified Model
of AGN (e.g. \citealt{antonucci93}). They are predominantly found
in quiescent galaxies and are often at the centres of groups or clusters, are
likely \foreignlanguage{british}{fuelled} by radiatively inefficient
accretion of hot gas, and would not be identified as AGNs at any other
wavelength (e.g.~\citealt{hardcastle07,smo08,smo09a,smo09b,hickox09}).
These AGNs are believed to significantly affect the evolution of their
host galaxies and perhaps of their environment on larger scales  (e.g.
\citealt{croton06,fabian06}). For example, mechanical energy transfer
via the radio jets may heat the intra-cluster/group gas and the
gaseous halo of the host galaxy, both of which are best traced via
X-rays (e.g. \citealt{fabian06}). This heating \textendash{} deemed
crucial in cosmological models of galaxy formation \textendash{} is
referred to as feedback; however,   it
is still not completely understood 
on group/cluster scales or on galaxy scales. In simulations of galaxy formation and
evolution, such radio-mode feedback plays a critical role in \foreignlanguage{british}{suppressing}
massive galaxy formation and reproducing various observed galaxy properties
(e.g. \citealt{croton06}). While this has been studied
for individual cases (e.g.~\citealt{worrall12}), robust observational
information relating to these feedback processes is limited by the
lack of statistically large samples of radio sources \citep{best06,merloni08,smo09a}.

To comprehensively examine the role of AGN feedback on their hosts
and environments, it is essential to obtain both radio and X-ray coverage
over large fields. This synergy allows a direct insight into different
heating mechanisms (thermal versus\ non-thermal) as a function of redshift,
source type, and galaxy location in large-scale structures. An example
of a previous radio survey with available X-ray coverage is the VLA-COSMOS
survey \citep{schinnerer07,smo14} with X-ray data from the XMM-Newton and Chandra observatories (\citealt{hasinger07,elvis09,civano12}; Civano et
al., \citetalias{civano_inprep}). This is a deep radio survey, reaching a sensitivity
of $\sim15\ \mu$Jy/beam at 20\,cm, but only covering a 2\,deg$^{2}$
field.
Another example is the Boötes survey (\citealt{hickox09,DeVries02}), which
covers a wider field (7\,deg$^{2}$) but is  shallower
(rms $\sim28\ \mu$Jy/beam at 20\,cm). To accurately measure
the evolution of the radio galaxy luminosity function, particularly
at the bright end, it is necessary to conduct very wide-field radio/X-ray
surveys while maintaining good sensitivity. For this purpose, we present
the first results from the pilot program of a large radio continuum
survey with the Australia Telescope Compact Array (ATCA) to cover
25\,deg$^{2}$ of the XXL survey field.

The XXL survey comprises the largest XMM-Newton project approved to
date (Pierre et al.\ 2015, \citetalias{pierre15} hereafter); it has provided
3~Ms of new data, and more than $6$~Ms when including archival data. The main goals
of this X-ray survey are to provide long-lasting legacy data for studies
of galaxy clusters and AGNs and to constrain the dark energy equation
of state using clusters of galaxies. Observations with XMM-Newton
are essentially complete;  a few fields are undergoing re-observation.
The survey covers an equatorial and a southern region of $\sim25$\,deg$^{2}$
each, down to a point-source sensitivity of $\sim5\times10^{-15}{\rm \ erg\, s^{-1}cm^{-2}}$
(0.5-2~keV). Several thousand spectroscopic redshifts of X-ray luminous
sources have been collected with the Anglo-Australian Telescope (Lidman et al.\ 2015, \citetalias{lidman15}) and photometric redshifts from existing optical/near-IR
(NIR) data will reach accuracies better than $\sim10\%$ (Fotopoulou et al.\ 2015, \citetalias{fotopoulou15}). See \citetalias{pierre15} for an overview of the
survey.

To provide the complementary radio data, we are undertaking a large
survey with ATCA to cover the full southern XXL field (hereafter XXL-S)
at 2.1\,GHz to an expected sensitivity of $\sim43$\,$\mu$Jy. These
observations will provide a unique resource of sensitive, complementary
radio and X-ray coverage over the widest field to date. This will
allow detailed studies of the role of radio-mode feedback on the formation
and evolution of massive galaxies and galaxy clusters.

Here we report the results of the pilot survey covering the central
6.5\,deg$^{2}$ of the field. We achieve a spatial resolution of
$\sim$4\arcsec\ and an average rms of $\sim$$50~\mu$Jy/beam. We
have identified 1389 radio sources in the field. Section 2 of this
paper details the ATCA observations, Section 3 discusses the data
reduction and imaging strategy, Section 4 presents the source identification
and catalogue, and Section 5 provides the summary.

\section{Observations, data reduction, and imaging}

\label{sec:obs}

\subsection{Observations}

We conducted 2.1 GHz observations with ATCA%
\footnote{\url{https://www.narrabri.atnf.csiro.au/observing/}} over 37 hours on 3-6 September 2012 in the 6A (6\,km) configuration
and over 15 hours on 25-26 November 2012 in the 1.5C (1.5\,km) configuration.
The observations were performed using the Compact Array Broadband
Backend (CABB; \citealt{wilson11}) correlator, which covers a full
2\,GHz bandwidth \foreignlanguage{british}{centred} on 2.1\,GHz
using a channel width of 1\,MHz. To cover the 6.5\,deg$^{2}$ pilot
field, 81 mosaic pointings were placed in  a layout such that the separation
between adjacent pointings in right ascension (RA) and declination
(DEC)  was two-thirds of the primary beam full-width at half-maximum (FWHM) at 2.1 GHz, i.e.
14.7\,arcmin. We used source 1934-638 as the primary calibrator,
and observed it during each observing run for 10 minutes on-source.
The flux density of 1934-638 was tied to the widely used absolute
flux density scale of \citet{baars77}, as described in \citet{reynolds94}.
Source 2333-528 was taken as the secondary calibrator and it was observed
for 2 minutes on-source every 32 minutes between observations of different
sets of pointings.

\subsection{Data reduction}

We used the Multichannel Image Reconstruction, Image Analysis and
Display (\textsc{miriad}) software package to reduce the data. In
the calibration step we used 16 frequency bins in the GPCAL (gain/phase/polarisation
calibration) task (i.e. bin width of 128~MHz). The 81 pointings were
individually reduced and imaged. Automated flagging was performed
using the MIRIAD task PGFLAG. This is based on AOFLAGGER and was
developed for LOFAR data sets, but has now been applied to various radio data from other instruments
\citep{offringa10,offringa12}. We found that the lowest frequency
sub-band, which was centred at 1.204~GHz, was significantly affected by radio
frequency interference (RFI). In the shortest baselines 80 -- 90\,percent
of the data in this sub-band was flagged, and even in the long 6\,km
baselines 60 -- 70\,percent of the data was affected by RFI. Consequently,
this sub-band was discarded completely.

\subsection{Imaging}

Wideband receivers such as CABB on the ATCA present new challenges
to radio imaging. The primary beam response, the synthesised beam,
and the flux density of most sources vary significantly with frequency
over the 2~GHz wide bandwidth. One method used to mitigate these issues
is to divide $(u,v)$ data into sub-bands and then to force similar beam sizes
with an appropriate ``robustness'' parameter \citep{briggs95}.
This approach was used to image VLA data spanning 2 -- 4 GHz \citep{condon12,novak15,smo15}.
We tested two imaging schemes, one where the $(u,v)$ data are not
divided into sub-bands and a second scheme where the 2048\,MHz CABB
band is divided into 256\,MHz sub-bands. We find that the effect
of bandwidth smearing is more significant in the first approach, thus
hereafter we focus and describe in detail only the second imaging
approach.

The calibrated data set of each pointing was split into eight sub-bands
of 256 MHz each. As the lowest sub-band is almost entirely contaminated
by RFI it was disregarded. Each sub-band was imaged with a robust
weighting chosen to match the beam sizes across the seven remaining sub-bands.
Multifrequency cleaning and self-calibration were performed for each
pointing and each sub-band. The task MFCLEAN was used to perform cleaning;
 the clean region was set to the inner 23$\arcmin$ region of the
image because a ``border'' is required for MFCLEAN. The cleaned region
extends beyond the 7\% primary beam response level at 2.1\,GHz, the
effective frequency of the observations, therefore encompassing the
full region of interest. We performed two iterations of phase self-calibration
to improve the images. The first self-calibration iteration was performed
with a model generated by cleaning to 10\,$\sigma$ (i.e. bright sources
only) and a second model generated by cleaning to 6\,$\sigma$.
The individual images were restored with the same beam, i.e. the average
beam of the 7 $\times$ 81 images,  $4.7\arcsec\times4.2\arcsec$.
The final sub-band combined mosaic was then obtained by using the
LINMOS task to create a noise-weighted mosaic of all 7 $\times$ 81
images. The different primary beam sizes for different sub-bands will
result in different effective frequencies for different positions
in the mosaic (see \citealt{condon12} for details). To take this
into account we generated a mosaic of effective frequencies using
the task LINMOS. The effective frequency of the central 5.2\,deg$^{2}$
region of the image has a median of 2.10\,GHz and a standard deviation
of only 0.07\,GHz.

The final mosaic, with an angular resolution of $4.7\arcsec\times4.2\arcsec$
and a pixel size of $\sim1''\times1''$, is shown in \f{fig:mosaic}
. We also overlaid the pixel flux distribution in $12\times12$ regions over the mosaic. The main image artefacts are stripes from imperfect
cleaning of sidelobes around bright sources and what are known as chequerboard
artefacts  around bright extended sources caused by missing short baselines.
The cleaning artefacts are caused by a variety of factors, including
imperfect antenna calibration and imperfect models for cleaning. For example,
MFCLEAN can only model the spectral energy distribution of a source
as a power law in log space ($S\propto\nu^{\alpha}$) and it uses
only point sources in the deconvolution. To improve on this, the new ATCA observations of the
full 25 deg$^{2}$ field have been performed with a more complete $uv$-coverage, and we are developing
a rigorous cleaning process to reduce sidelobes around the bright
sources and peeling (e.g.\ Intema et al. 2009) of off-axis bright
sources to reduce the artefacts in neighbouring pointings. This will
allow us to minimise the rms of the image and to identify and catalogue
sources down to a $5\sigma$ level with a high level of completeness. In \f{fig:visibility} \ we show the mosaic sensitivity function,
i.e.\ the total area covered at a given rms. The mosaic reaches a median
rms of $50~\mu$Jy/beam.

\begin{figure*}
\includegraphics[bb=0bp 150bp 700bp 800bp,clip,scale=0.9]{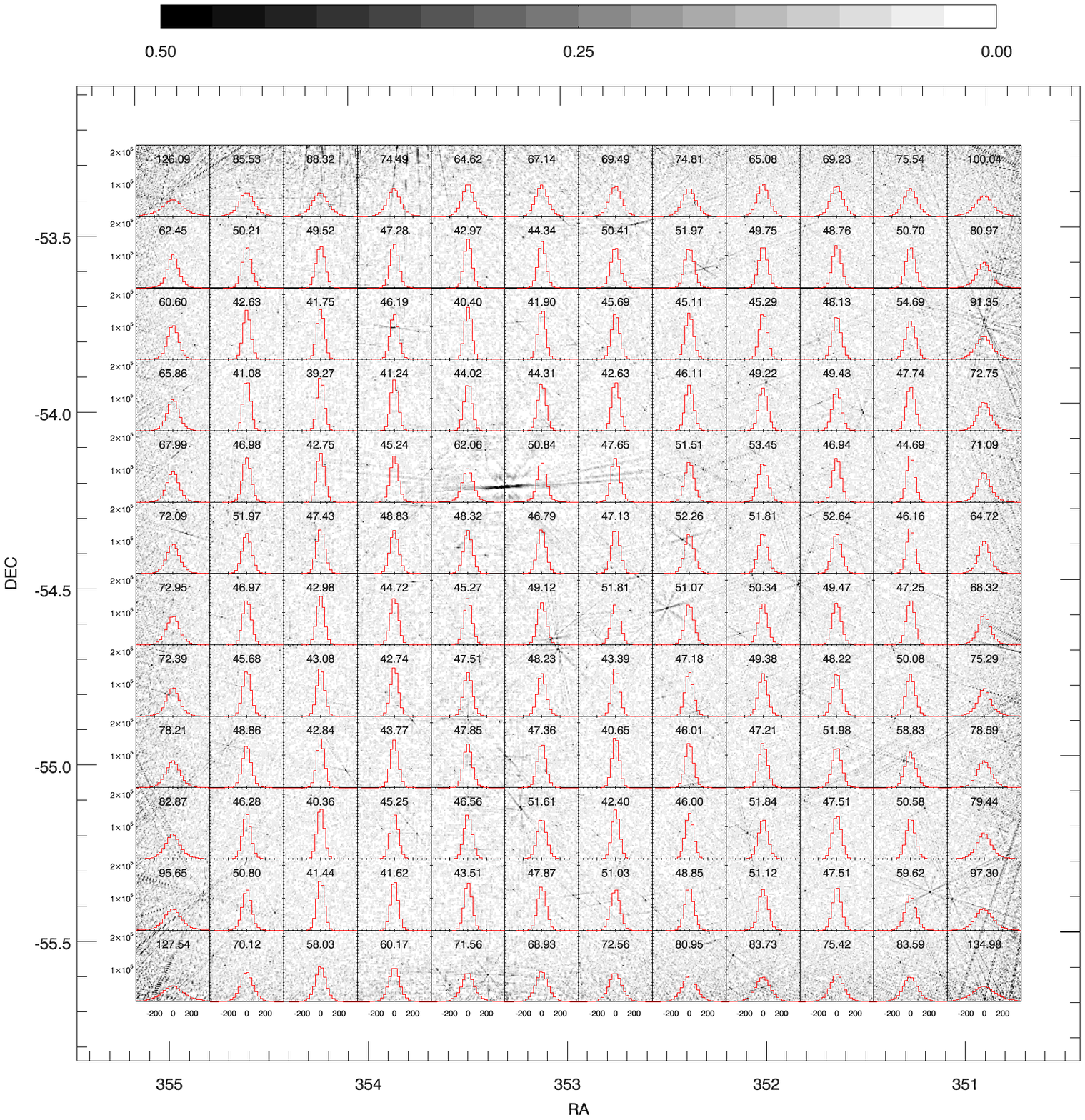}
\protect\protect\caption{Inverted greyscale of our XXL-S Pilot Survey mosaic at 2.1~GHz. The
greyscale colourbar in units of mJy/beam is shown at the top. Overlaid
are histograms of pixel flux distributions for different regions of
the mosaic. The standard deviation for each region is indicated in
each panel (in units of $\mu$Jy/beam). The x-axis of the histograms
shows flux density in $\mu$Jy/beam and the y-axis shows counts. The
outer 500 pixels of the map have been excluded here since they are
not considered in the source-finding process. \label{fig:mosaic}}
\end{figure*}

\begin{figure}
\includegraphics[bb=0bp 230bp 600bp 600bp,clip,width=1\columnwidth]{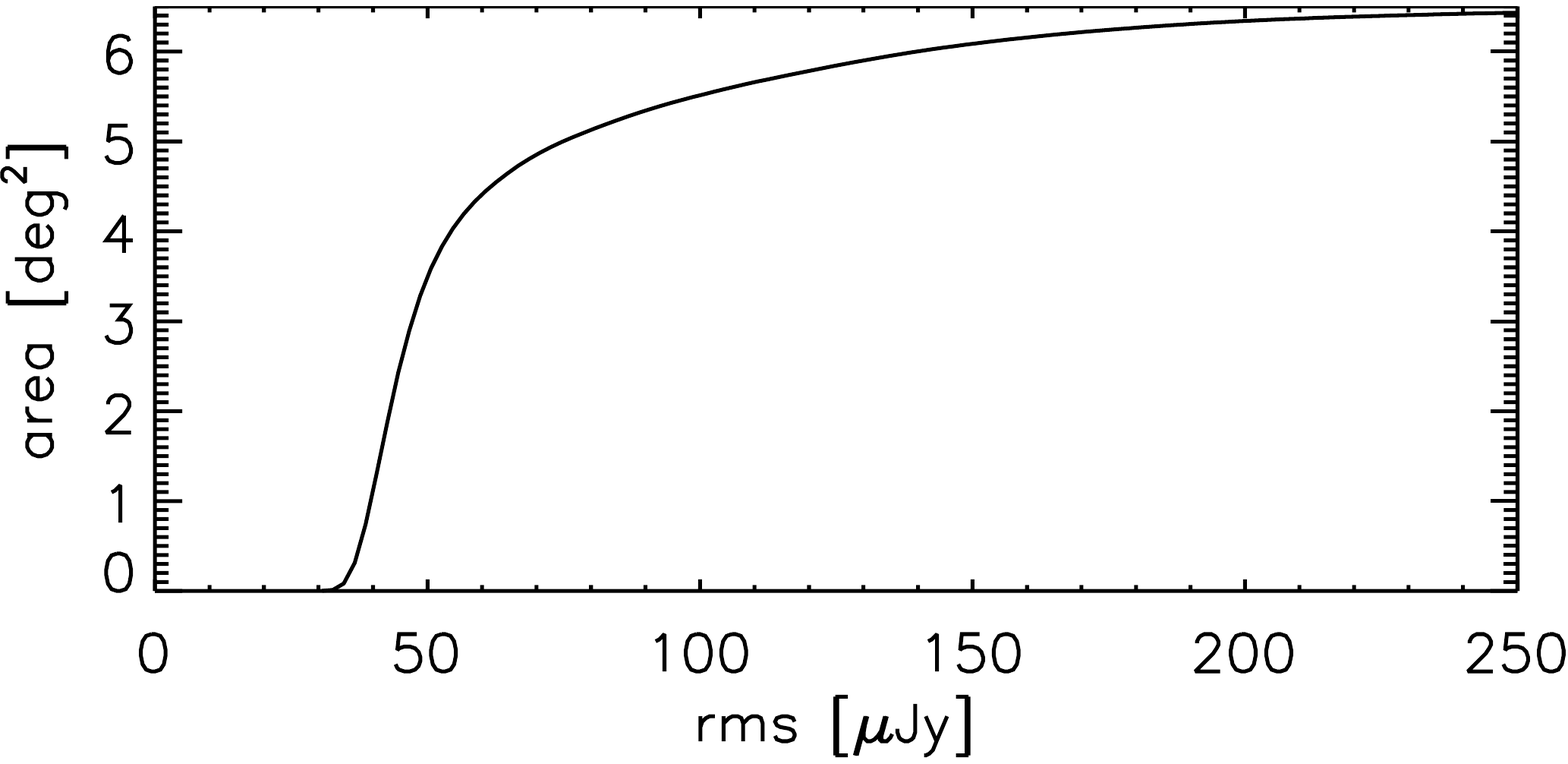}
\protect\protect\caption{The mosaic sensitivity function showing the total area with a noise
level below a given rms. \label{fig:visibility}}
\end{figure}

\section{ Source catalogue\label{sec:catalog}}

\subsection{Noise image\label{sub:Noise-image}}

In order to select a sample of radio sources above a given threshold,
defined in terms of local signal-to-noise ratio (S/N), we derived
a noise image using the task RMSD in the Astronomical Image Processing System (\textsc{aips}). The rms was calculated for
each pixel based on the data in the box surrounding the pixel. The
size of the box was chosen to be 60 pixels (i.e. $60''\times60''$).
The rms was calculated on a multi-iteration basis, which excludes the points
 that are more than 3 times the rms derived from the previous iteration until convergence
is reached. The noise distribution is not entirely Gaussian owing to
the presence of residual sidelobes around the many bright radio sources
in the field. The median rms is 50 $\mu$Jy/beam. About 3\% of the
noise values are higher than 200 $\mu$Jy and are located around the
brightest sources in the mosaic. Using the total intensity and noise
images we produced a S/N map. The minimum and maximum values of this
S/N map are -7.5 and 1006, respectively.

\subsection{Source detections\label{sub:Source-detections}}

To extract a catalogue of sources we followed the procedure already
applied to the COSMOS field and tested by \citet{schinnerer07,schinnerer10}
and \citet{smo14}. We first ran the \textsc{aips} task SAD on the S/N map
to derive a catalogue of radio components. At this stage, we used
a threshold of 4.7 in peak S/N. For each selected component, the peak
surface brightness (hereafter peak flux), the total flux density,
the position, and the size were estimated using a Gaussian fit. However,
for faint components the optimal estimate of the peak flux and of
the component position were obtained by a non-parametric interpolation
of the pixel values around the fitted position using the task MAXFIT
in \textsc{aips} (as in \citealt{schinnerer07}). Only the components with
a S/N (derived as the ratio between the MAXFIT peak brightness and
the local noise at the position of the MAXFIT peak) $\geq$5 are included in the catalogue. Around the brightest sources ( $\sim50$~mJy/beam)
residual sidelobes can be mistakenly identified as real components
by SAD. These regions were inspected by eye to remove sidelobe spikes.
We further excluded all data less than 500 pixels (i.e. $500''$) from
the outer edge of the noise map where the rms is too large for reliable
source-finding.

\begin{figure}
\includegraphics[bb=20bp 0bp 550bp 780bp,clip,width=0.75\columnwidth,angle=90]{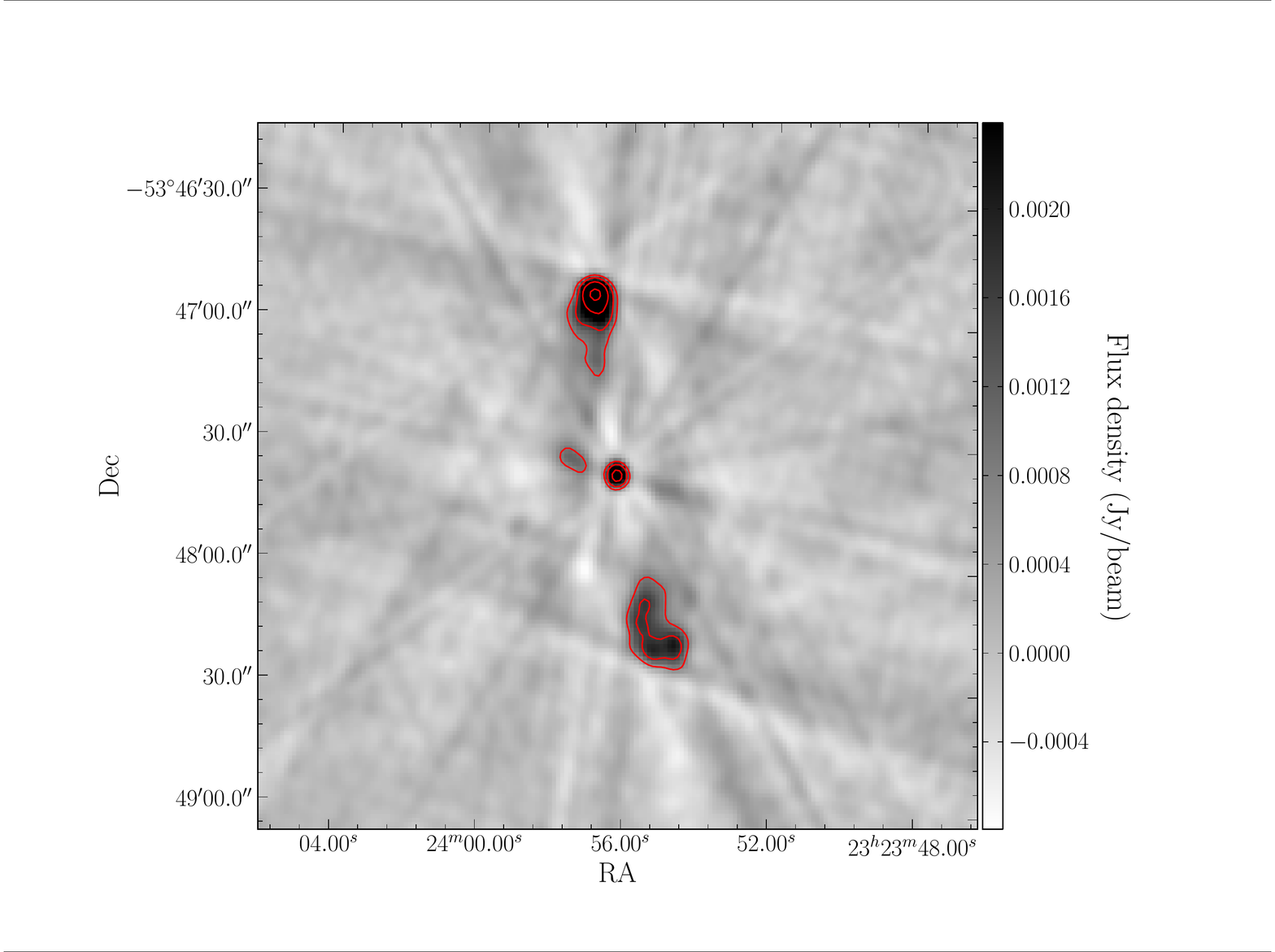}
\protect\protect\caption{Example of a multicomponent source. The greyscale 2.1\,GHz
image is shown in the background; the red radio flux contours are shown
at 0.8, 1.6, 3.2, 6.4, 12.8\,mJy/beam. \label{fig:galaxy}}
\end{figure}

Some of the components identified by SAD clearly belong to a single
radio source,  for example the two lobes of FRII radio sources \citep{fanaroff74}
or extended emission associated with compact components. We visually
checked all these cases and compared them with available deep optical images
(from the Blanco Cosmology Telescope (BCS) and also DECcam and VISTA-NIR
images; see \citetalias{pierre15} and references therein) in order to
associate multiple components with a single radio source.  If an optical
counterpart can be found, this often helps to determine whether
two radio components are associated with one central galaxy. An example
of such a multicomponent system is shown in Figure \ref{fig:galaxy}. 

Other automatic methods that are based on a statistical approach could
also be used to  identify multiple radio components
belonging to a single source; however, our experience has taught us that a classification
based on the radio morphology of the component supported by deep optical/NIR
images is more effective. This task is clearly time consuming for large
surveys, but it will be adopted for the whole survey.

The final catalogue lists 1389 radio sources with S/N$\geq$5. Of these, 77 are multiple, i.e. fitted with at least two separate
components. For the multicomponent sources we recalculated the flux
using the \textsc{aips} task TVSTAT which allows the manual definition
of the integration bounds, which we set to 2$\sigma$ contours. In
the catalogue this value is  reported for the total flux, the
peak flux is set to -99, and the sources are flagged as multicomponent
sources (MULT=1).

\begin{figure}
\includegraphics[bb=55bp 0bp 650bp 480bp,clip,scale=0.45]{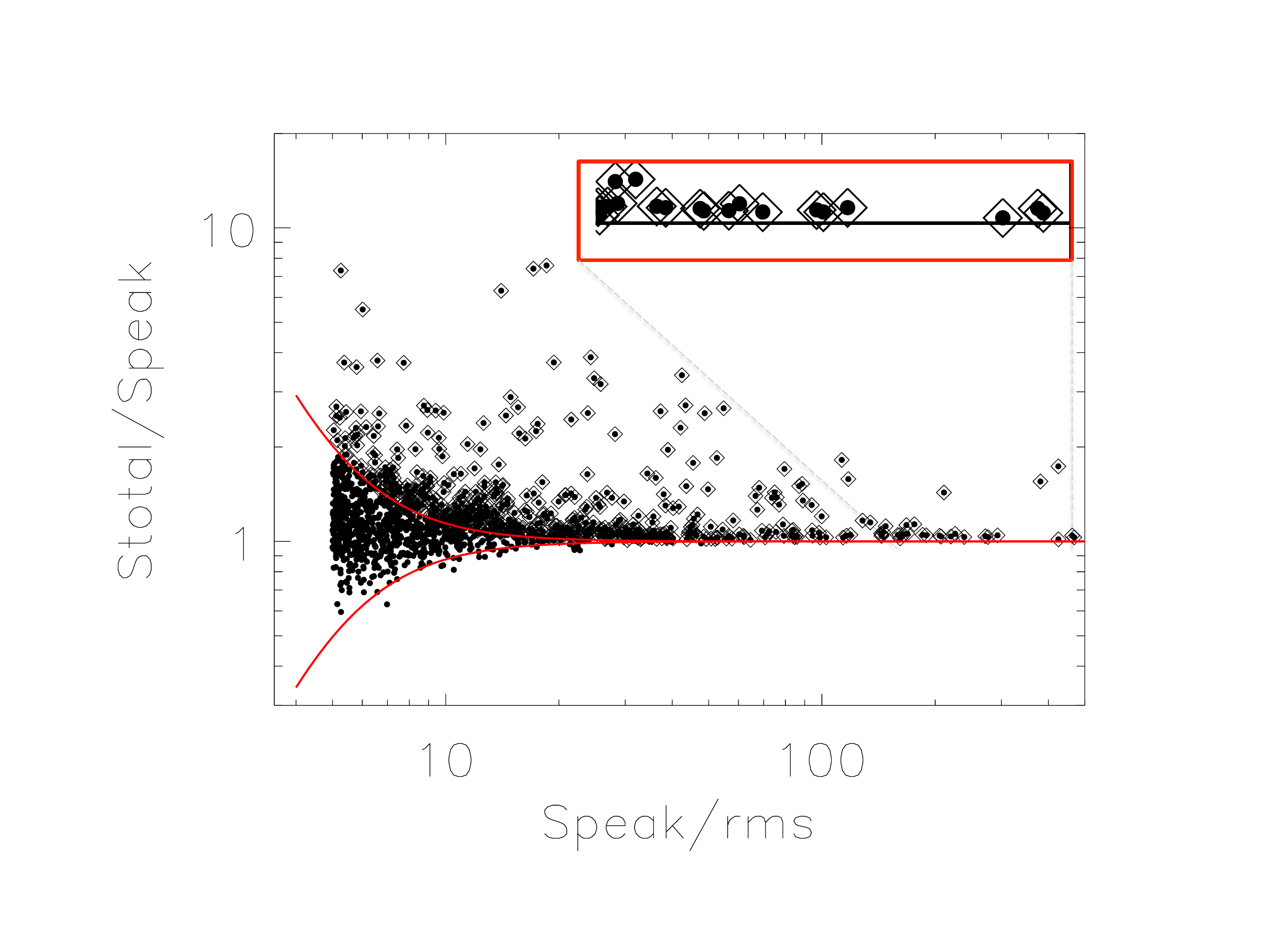}
\protect\protect\caption{Ratio of total to peak flux of sources identified in the 2\,GHz
mosaic versus the peak flux to rms ratio. Also shown are the boundary
encompassing 90\% of the sources (lower line) and its reflection
about the ratio=1 value (upper line), above which the sources are considered
to be resolved (marked by open symbols). The inset shows a magnified
view of the high S$_{{\rm peak}}$ region (roughly $180<S_{{\rm peak}}/\mbox{{\rm rms}}<450$)
where the vertical offset from 1.0 due to the bandwidth smearing is
clearly visible (see Section \ref{sub:Source-detections}). \label{fig:StSp} }
\end{figure}

Figure \ref{fig:StSp} shows the ratio of total flux (S$_{\mathrm{T}}$)
to peak flux (S$_{\mathrm{P}}$) versus the S/N of the catalogued
single-component sources. As described in \citet{bondi03}, this plot
can be used to distinguish between point (unresolved) sources and
extended (resolved) sources. In this figure, the vertical offset from
1.0 at the very bright end is caused by bandwidth smearing in the
mosaic (see insert in Figure \ref{fig:StSp} and \citealt{bondi08}
for details). Bandwidth smearing affects the fluxes in such a way
that the peak flux  decreases by a certain amount, while the total
flux remains the same. Based on the data shown in \f{fig:StSp} , we estimated a 4\% bandwidth smearing effect, and thus corrected all
peak fluxes by multiplying them by 1.04. Accordingly, following \citet{bondi03,bondi08}
we fitted a lower envelope to the data, which contains 90\% of the sources
with S$_{\mathrm{T}}<\mathrm{S}_{\mathrm{P}}$ (with S$_{\mathrm{P}}$
corrected for the 4\% bandwidth smearing), and mirrored it above S$_{\mathrm{T}}/\mathrm{S_{P}}=1.00$. The upper envelope, above which sources are considered resolved,
is then given by 
\begin{equation}
\mathrm{S_{T}/S_{P}}=1+\frac{90}{(\mathrm{S_{P}}/rms)^{2.35}}
.\end{equation}

In total we find 228 single-component sources to be resolved (in addition
to the 77 multicomponent sources). The resolved sources are flagged
in the catalogue by RES = 1. For the unresolved sources, the total
flux density is set equal to the peak flux density and the angular
size is set equal to zero in the catalogue.

Finally, we calculated the uncertainties $\sigma_{S_{T}}$ and $\sigma_{S_{P}}$
on the total (S$_{T}$) and peak (S$_{P}$) fluxes using the method
described in detail in Condon (1997, see also Schinnerer et al. 2007),
$\sigma_{S_{P}}^{2}$ = 2 $S_{P}^{2}$ / $\rho^{2}$ and $\sigma_{S_{T}}^{2}$
= 2 $S_{T}^{2}$ / $\rho^{2}$, where $\rho$ is the signal to noise
ratio given by

\[
\rho^{2}=\frac{\theta_{M}\theta_{m}}{4\theta_{N}^{2}}\left[1+\left(\frac{\theta_{N}}{\theta_{M}}\right)^{2}\right]^{3/2}\left[1+\left(\frac{\theta_{N}}{\theta_{m}}\right)^{2}\right]^{3/2}\frac{S_{P}^{2}}{\sigma_{map}^{2}}
\]

where $\theta_{M}$ and $\theta_{m}$ are the fitted FWHMs of the
major and minor axes, $\sigma_{map}$ is the noise of the image, and
$\theta_{N}$ is the FWHM of the synthesised beam.

The positional errors are estimated using

$\sigma_{RA}^{2}$ = $\epsilon_{RA}^{2}$ + $\sigma_{xo}^{2}$ sin$^{2}$(PA)+$\sigma_{yo}^{2}$cos$^{2}$(PA) and

$\sigma_{DEC}^{2}$ = $\epsilon_{DEC}^{2}$ + $\sigma_{xo}^{2}$ cos$^{2}$(PA)+$\sigma_{yo}^{2}$sin$^{2}$(PA),

where PA is the positional angle of the major axis, $\epsilon_{RA}^{2}$
and $\epsilon_{DEC}^{2}$ are the calibration errors, while $\sigma_{xo}$
and $\sigma_{yo}$ are $\theta_{M}^{2}$(4 ln 2)$\rho^{2}$ and $\theta_{m}^{2}$/(4
ln 2)$\rho^{2}$, respectively. Calibration terms $\epsilon_{RA}^{2}$
and $\epsilon_{DEC}^{2}$ must be estimated from comparison
with external data with better positional accuracy than the one tested.
We calculated our calibration terms from the comparison between the
position of single-component XXL ATCA radio sources with S/N$>$10
and their optical counterpart (62 sources in total).  The mean values
and standard deviations found from this comparison are $\Delta$RA~=~0.24~$\pm$~0.34~arcsec
and $\Delta$DEC~=~$-$0.12~$\pm$~0.22~arcsec. These position
offsets are barely significant (6 and 4 times the error on the mean,
respectively). Given the size of the beam and the error associated
with the position of the bulk of weaker sources (comparable to or larger
than the previous offsets) at this stage we assume no significant
offset between the radio and optical frames. Once the full XXL-S field
has been covered, we plan to investigate in greater detail any possible systematic
and/or position-dependent offsets between the radio and optical positions.
Therefore, in the error budget we assume a calibration error $\epsilon_{RA}^{2}$
= 0.34 arcsec in RA and $\epsilon_{DEC}^{2}$ = 0.22 arcsec in DEC.

An example page of the final source catalogue is shown in Table \ref{tab:5cat}. The full catalogue is available as a queryable database table `XXL\_ATCA\_15'
via the XXL Master Catalogue browser\footnote{\url{http://cosmosdb.iasf-milano.inaf.it/XXL}}. A copy will also be deposited at the Centre de Donn\'ees astronomiques de Strasbourg (CDS)\footnote{The full source catalogue is available in electronic form
at the CDS via anonymous ftp to cdsarc.u-strasbg.fr (130.79.128.5)
or via http://cdsweb.u-strasbg.fr/cgi-bin/qcat?J/A+A/}. \\

\begin{table*}
\protect\protect\caption{Sample catalogue page. Column (1) gives the ID of the source, Cols.
(2) to (5) list the Right Ascension (RA) and Declination (DEC) in
J2000 epoch, both in units of decimal degrees and sexagesimal (hours,
minutes, and seconds for RA and degrees, arcminutes, and arcseconds
for DEC); Cols. (6) and (7) give the positional errors on RA and
DEC in arcseconds, Cols. (8) to (11) give the peak surface brightness
and total flux with their errors in units of mJy/beam and mJy respectively.
Column (12) gives the noise level in mJy at the source position. The
deconvolved major axis, minor axis, and position angle are given in Cols.
(13) to (15). Beam deconvolution is automatically output by the \textsc{aips}
tasks SAD/JMFIT. Column (16) gives the RES flag as 1 (0) if the
source is resolved (unresolved). Column (17) gives the MULT flag as
1 (0) if the source is a multicomponent (single) source. The last
column gives the S/N.\label{tab:5cat} }

\resizebox{18.5cm}{!}{ %
\begin{tabular}{|c|c|c|c|c|c|c|c|c|c|c|c|c|c|c|c|c|c|}
\hline 
ID  & RA$_{\mathrm{J2000}}$  & DEC$_{\mathrm{J2000}}$  & RA$_{\mathrm{J2000}}$  & DEC$_{\mathrm{J2000}}$  & $\Delta$RA  & $\Delta$DEC  & S$_{\mathrm{peak}}$  & S$_{\mathrm{peak-err}}$  & S$_{\mathrm{total}}$  & S$_{\mathrm{total-err}}$  & rms  & MAJAX  & MINAX  & PA  & RES  & MULT  & S/N \tabularnewline
 & {\tiny{}{[}$^{\circ}${]}}  & {\tiny{}{[}$^{\circ}${]}}  & {[}h m s{]}  & {[}d m s{]}  & {[}$arcsec${]}  & {[}$arcsec${]}  & {[}mJy/beam{]}  & {[}mJy/beam{]}  & {[}mJy{]}  & {[}mJy{]}  & {[}mJy{]}  & {[}$arcsec${]}  & {[}$arcsec${]}  & {[}$deg${]}  &  &  & \tabularnewline
\hline 
XXL-ATCA J232300.7-551110  & 350.7529585  & -55.1863031  & 23 23 00.710  & -55 11 10.691  & 0.35  & 0.26  & 2.510  & 0.223  & 4.507  & 0.400  & 0.246  & 6.08  & 1.85  & 103.10  & 1  & 0  & 9.80 \tabularnewline
XXL-ATCA J232304.4-541907  & 350.7684192  & -54.3186713  & 23 23 04.421  & -54 19 07.217  & 0.38  & 0.35  & 1.044  & 0.159  & 1.044  & 0.159  & 0.164  & 0.00  & 0.00  & 0.00  & 0  & 0  & 6.13 \tabularnewline
XXL-ATCA J232305.4-543631  & 350.7726871  & -54.6087976  & 23 23 05.445  & -54 36 31.671  & 0.34  & 0.22  & 9.744  & 0.204  & 10.581  & 0.222  & 0.207  & 1.67  & 0.99  & 92.10  & 1  & 0  & 45.32 \tabularnewline
XXL-ATCA J232306.9-534612  & 350.7791203  & -53.7700122  & 23 23 06.989  & -53 46 12.044  & 0.41  & 0.27  & 0.869  & 0.130  & 0.869  & 0.130  & 0.134  & 0.00  & 0.00  & 0.00  & 0  & 0  & 6.25 \tabularnewline
XXL-ATCA J232308.7-540346  & 350.7864525  & -54.0630222  & 23 23 08.749  & -54 03 46.880  & 0.34  & 0.22  & 16.435  & 0.227  & 19.943  & 0.275  & 0.235  & 3.47  & 0.00  & 55.30  & 1  & 0  & 67.37 \tabularnewline
\hline 
\end{tabular}} 
\end{table*}

\section{Comparison with other radio data}

\subsection{Flux comparison and spectral indices\label{sub:Flux-comparison}}

\begin{figure}
\includegraphics[bb=30bp 120bp 580bp 700bp,clip,scale=0.45]{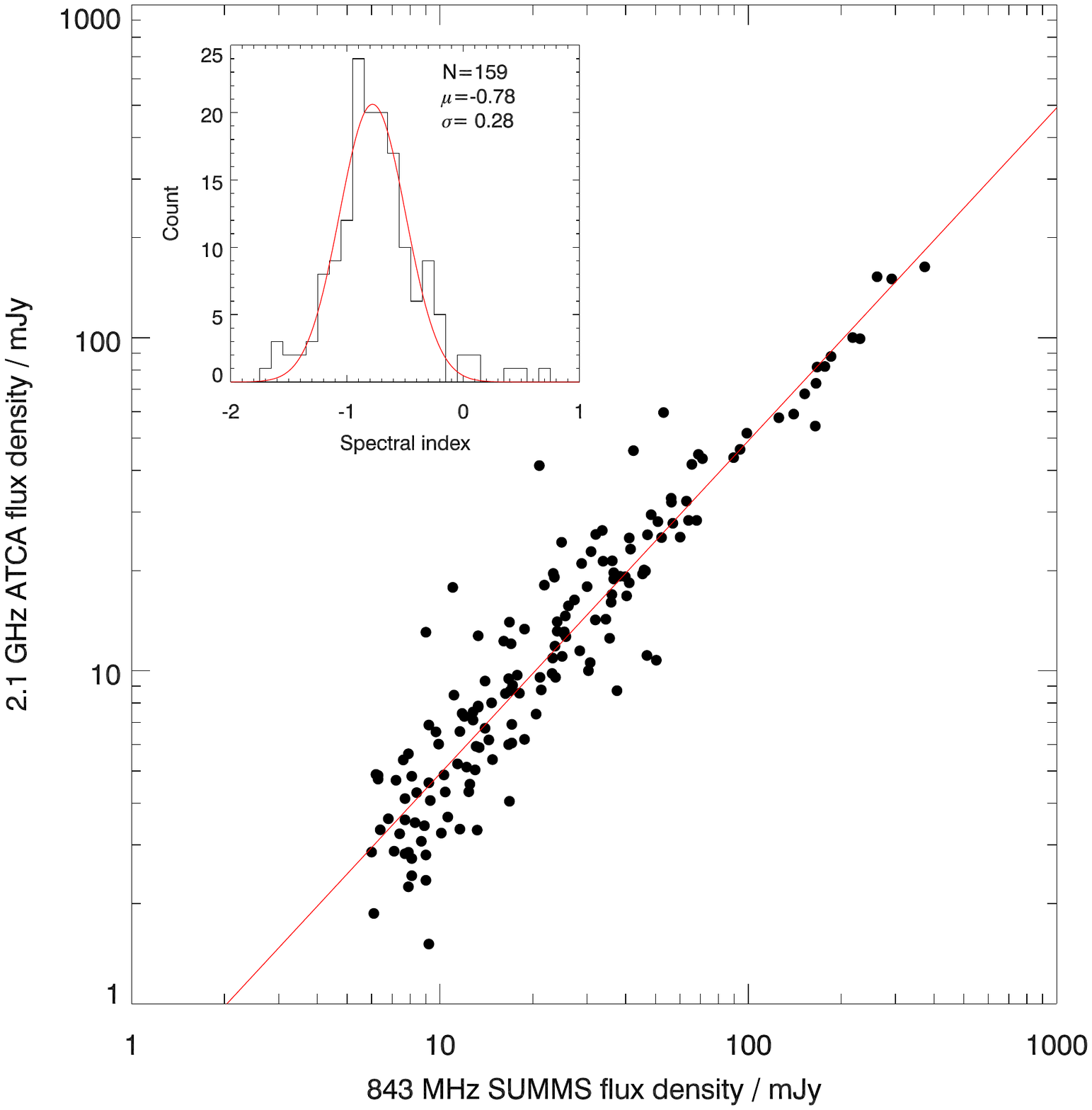}
\protect\protect\caption{Comparison between SUMSS 843MHz and ATCA-XXL 2.1 GHz fluxes for the
XXL-S Pilot area based on 159 sources detected in the shallower (rms
$\sim1.25$\,mJy/beam) SUMSS survey. The inset shows the spectral
index distribution for the sources with a mean of -0.78 and a standard
deviation of 0.28. The solid line corresponds to a spectral index
of $-0.78$. \label{fig:fluxcomp}}
\end{figure}

In the radio regime, the area of XXL-S is covered (with sufficient
sensitivity for comparison) by the Sydney University Molonglo Sky
Survey (SUMSS) survey%
\footnote{\url{http://www.physics.usyd.edu.au/sifa/Main/SUMSS}%
} \citep{bock99,mauch03}. Completed in 2007, SUMSS was conducted at
843~MHz and an angular resolution of $45\arcsec\times45\arcsec\mathrm{cosec}|\delta|$,
and it covers almost the whole sky south of Declination -30 degrees.
The SUMSS source catalogue \citep{mauch03} lists sources brighter
than 6~mJy/beam in peak flux. Cross-correlating our 2.1~GHz source
catalogue with the SUMSS survey catalogue using a radius of $15\arcsec$,
we find 159 matches after excluding faint multiple sources blended
in the SUMSS data. A comparison of the 2.1\,GHz ATCA and 843\,MHz
SUMSS fluxes for these sources is shown in Figure \ref{fig:fluxcomp}.
Also shown is the spectral index distribution of the sources; 
the spectral index ($\alpha$) is defined such that S$_{\nu}\propto\nu^{\alpha}$,
where S$_{\nu}$ is the flux at frequency $\nu$. A Gaussian fit to
the spectral index distribution yields a mean of -0.78 and a standard
deviation of 0.28. These values are consistent with expectations (e.g.
\citealt{kimball08}).

\subsection{XXL ATCA source counts}

In this section we report the 2.1\,GHz radio source counts obtained
in our ATCA observation in the XXL-S field. The completeness of the
sample will be tested with the appropriate simulation (see \citealt{bondi03}
for the description of the method) when the full radio data set becomes
available over the entire XXL-S field. Therefore, in order to reduce
problems with possible spurious sources near the flux limit and the effects
of incompleteness, we construct the 2.1~GHz ATCA radio source counts
considering only the 1003 sources in the pilot survey with a flux
density greater than 0.35~mJy, corresponding to S/N $\gtrsim$ 7.0
(assuming an average rms of 0.05~mJy, see section \ref{sub:Noise-image}).

The source counts from our pilot sample are summarised in Table \ref{tab:counts}
where, for each flux density bin, we report the minimum and mean flux
density, the observed number of sources, the differential source density
$n=dN/dS$ (in sr$^{-1}$ Jy$^{-1}$), the normalised differential
counts $nS^{2.5}$ (in sr$^{-1}$ Jy$^{1.5}$) with the estimated
Poisson error (as $n^{1/2}S^{2.5}$), and the integrated counts $N(>S)$
(in deg$^{-2}$).

\begin{center}
\begin{table*}
\begin{centering}
\protect\protect\caption{The 2.1 GHz radio source counts for the pilot sample in the XXL-S
survey. Column (1) gives the minimum flux density in mJy, Col. (2) gives
the mean flux density in mJy. In Col. (3) the observed number of
sources is stated. Column (4) gives the differential source density
($n$) in units of sr$^{-1}$~Jy$^{-1}$. Normalised differential
counts and their Poisson errors are given in Col. (5). The number
of sources in deg$^{-2}$ with Poisson error is given in Col. (6).
\label{tab:counts}}

\par\end{centering}

\centering{}%
\begin{tabular}{crccrc}
\hline 
$S$  & $<S>$  & N  & $n=dN/dS$  & $nS^{2.5}$  & $N(>S)$ \tabularnewline
{{[}mJy{]}}  & {{[}mJy{]}}  &  & {{[}sr$^{-1}$ Jy$^{-1}${]}}  & {{[}sr$^{-1}$ Jy$^{1.5}${]}}  & {{[}deg$^{-2}${]}} \tabularnewline
\hline 
0.35  & 0.46  & 304  & 1.09 $\times$10$^{9}$  & 4.8 $\pm$ 0.3  & 197.4 $\pm$ 6.5 \tabularnewline
0.59  & 0.78  & 216  & 3.15 $\times$10$^{8}$  & 5.3 $\pm$ 0.4  & 116.3 $\pm$ 4.4 \tabularnewline
1.01  & 1.32  & 139  & 1.06 $\times$10$^{8}$  & 6.7 $\pm$ 0.6  & 76.3 $\pm$ 3.5 \tabularnewline
1.71  & 2.24  & 128  & 5.43 $\times$10$^{7}$  & 12.9 $\pm$ 1.1  & 53.5 $\pm$ 2.9 \tabularnewline
2.92  & 3.81  & 60  & 1.48 $\times$10$^{7}$  & 13.3 $\pm$ 1.7  & 33.6 $\pm$ 2.3 \tabularnewline
4.96  & 6.48  & 44  & 6.39 $\times$10$^{6}$  & 21.6 $\pm$ 3.3  & 24.4 $\pm$ 1.9 \tabularnewline
8.44  & 11.01  & 45  & 3.82 $\times$10$^{6}$  & 48.7 $\pm$ 7.3  & 17.6 $\pm$ 1.6 \tabularnewline
14.36  & 18.72  & 30  & 1.50 $\times$10$^{6}$  & 72 $\pm$ 13 & 10.7 $\pm$ 1.3 \tabularnewline
24.41  & 31.83  & 15  & 4.41 $\times$10$^{5}$  & 80 $\pm$ 21  & 6.1 $\pm$ 1.0 \tabularnewline
41.50  & 54.12  & 13  & 2.25 $\times$10$^{5}$  & 153 $\pm$ 42  & 3.8 $\pm$ 0.8 \tabularnewline
70.56  & 91.99  & 7  & 7.12 $\times$10$^{4}$  & 183$\pm$ 69  & 1.8 $\pm$ 0.5\tabularnewline
\hline 
\end{tabular}
\end{table*}

\par\end{center}

\begin{figure*}
\centering{}\includegraphics[bb=65bp 400bp 570bp 750bp,clip,scale=0.7]{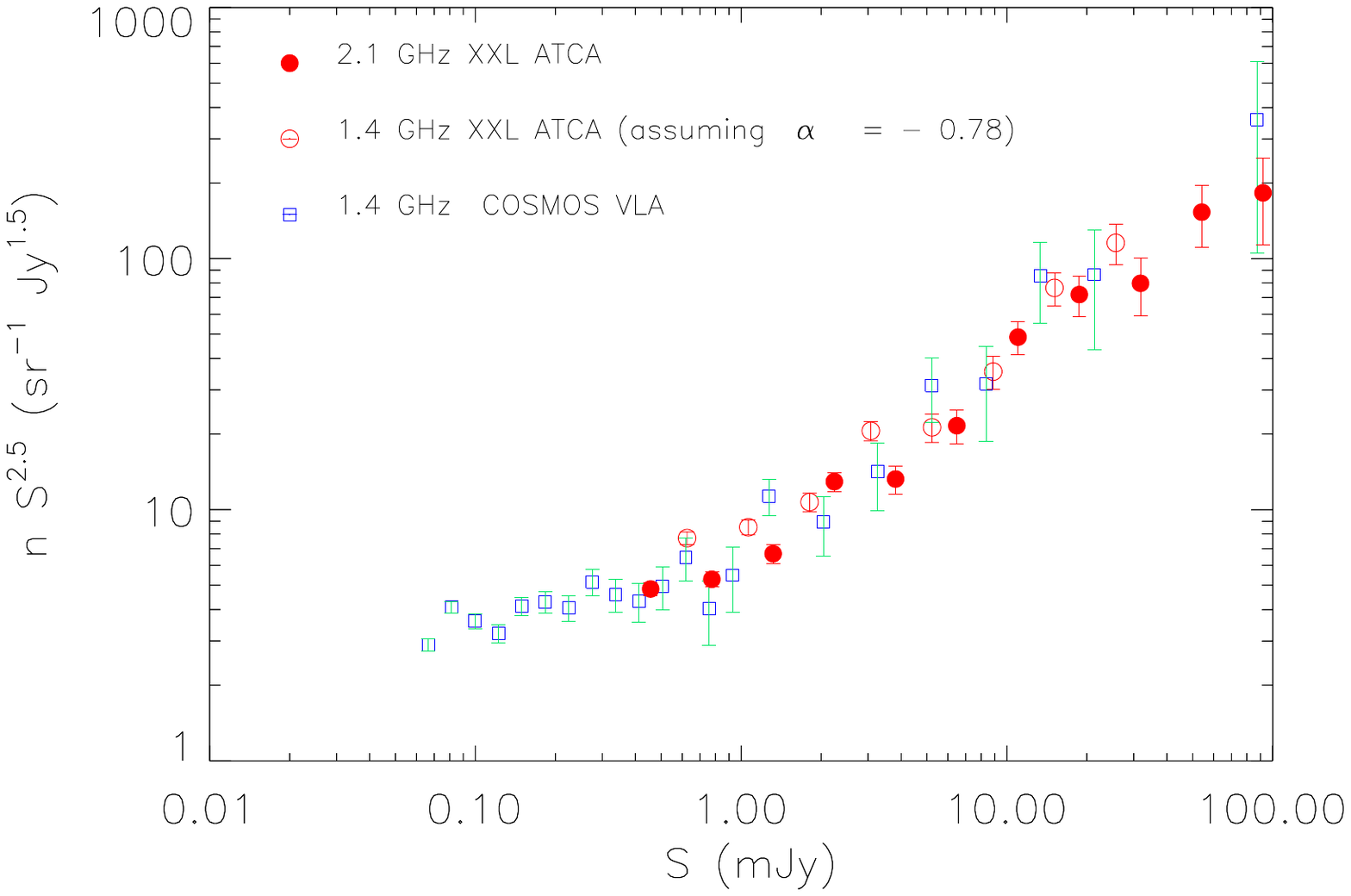}
\protect\protect\caption{ Normalised differential source counts for the 2.1~GHz ATCA XXL-S
survey (red filled dots), estimated 1.4\,GHz ATCA XXL-S counts (assuming
a spectral index of -0.78; red open dots) and for the COSMOS VLA 1.4
GHz survey (blue open squares) \label{fig:counts}}
\end{figure*}

The normalised differential counts $nS^{2.5}$ are plotted in Figure
\ref{fig:counts} where, for comparison, the differential source counts
obtained from the COSMOS VLA survey at 1.4~GHz (\citealt{bondi08})
are also plotted. For ease of comparison, the estimated ATCA source
counts at 1.4\,GHz are also plotted assuming a spectral index of
-0.78 for all sources (i.e. the average value derived in section \ref{sub:Flux-comparison}).
We extrapolate the $7\sigma$ cutoff of 0.35 mJy at 2.1 GHz to 1.4
GHz using the average spectral index of $\alpha=-0.78$, i.e.\ the
1.4 GHz extrapolated counts are cut off at 0.48 mJy. As
shown in Figure \ref{fig:counts}, our counts are in good agreement
with the COSMOS survey counts over the full flux density range sampled
by our survey ($\sim$ 0.35 - 100~mJy), indicating that the survey
is relatively complete, at least down to $7\sigma$.

\section{Summary\label{sec:summary}}

We have presented a 2.1\,GHz ATCA continuum map of the central 6.5\,deg$^{2}$
of the southern XXL field. This pilot survey is part of a larger observing
program currently underway to map the full 25\,deg$^{2}$ XXL-S field.
This will provide sensitive, wide-field continuum data with complementary
X-ray coverage allowing detailed studies of the heating mechanisms
of radio AGN and galaxy clusters.

Our final continuum map has an angular resolution of $4.7\arcsec\times4.2\arcsec$
and an rms of $\sim$50\,$\mu$Jy/beam. There are 1389 radio sources above
5\,$\sigma$  identified in the map, 77 of which consist of multiple
components. The differential source counts are consistent with those
found in the COSMOS-VLA survey at 1.4\,GHz. By comparing the 2.1\,GHz ATCA fluxes with the
843\,MHz SUMSS survey, we find an average spectral index for these
sources of -0.78 with a scatter of 0.28, which is consistent with previous
findings.

The ATCA observations of the remainder of the XXL-S field have been completed
and data reduction is underway. The focus is now on developing a rigorous
cleaning process to reduce sidelobes around the bright continuum sources,
to \foreignlanguage{british}{minimise} the rms of the image, and to
identify and catalogue sources down to a $\sim5$\,$\sigma$ level
with good completeness. 
\begin{acknowledgements}
XXL is an international project based on an XMM Very Large \foreignlanguage{british}{P}rogramme
surveying two 25~deg$^{2}$ extragalactic fields at a depth of $\sim5\times10^{-15}$~erg~cm$^{-2}$~s$^{-1}$
in the {[}0.5-2{]}~keV band for point-like sources. The XXL website
is \texttt{http://irfu.cea.fr/xxl}. Multiband information and spectroscopic
follow-up of the X-ray sources are obtained through a number of survey
programmes, summarised at \texttt{http://xxlmultiwave.pbworks.com/}.

The Australia Telescope Compact Array is part of the Australia Telescope
National Facility which is funded by the Commonwealth of Australia
for operation as a National Facility managed by CSIRO. This research
was funded by the European Union's Seventh Framework program under
grant agreement 333654 (CIG, `AGN feedback'). VS, JD and MN acknowledge
funding from the European Union's Seventh Framework program under
grant agreement 337595 (ERC Starting Grant, `CoSMass'). VS acknowledges
funding from the Australian Group of Eight European Fellowships 2013. FP acknowledges support from the BMBF/DLR grant 50 OR 1117, the DFG grant RE 1462-6 and the DFG Transregio Programme TR33. MNB and M Birkinshaw acknowledge funding from the UK's STFC. HR acknowledges support from the ERC Advanced Investigator program NewClusters 321271.
\end{acknowledgements}

 \bibliographystyle{aa}
\bibliography{atca-xxl}

\end{document}